\documentclass[letterpaper]{article} 
\usepackage{aaai24}  
\usepackage{times}  
\usepackage{helvet}  
\usepackage{courier}  
\usepackage[hyphens]{url}  
\usepackage{graphicx} 
\usepackage{natbib}  
\usepackage{caption} 
\begin{document}

\title{
 Remote Possibilities:  Where There Is a WIL, Is There a Way? \\AI Education for Remote Learners in a New Era of Work-Integrated-Learning}
\author{Derek Jacoby, Saiph Savage and Yvonne Coady
\\ Computer Science University of Victoria
\\ Khoury College of Computer Science Northeastern University}

\author {
    Derek Jacoby\textsuperscript{\rm 1},
    Saiph Savage\textsuperscript{\rm 2},
    Yvonne Coady\textsuperscript{\rm 1, \rm 2}
}
\affiliations {
    \textsuperscript{\rm 1}University of Victoria, derekja@uvic.ca\\
    \textsuperscript{\rm 2}Northeastern University 
}
\maketitle

\begin{abstract}
\begin{quote}

{
Increasing diversity in educational settings is challenging in part due to the lack of access to resources for non-traditional learners in remote communities.  Post-pandemic platforms designed specifically for remote and hybrid learning---supporting team-based collaboration online---are positioned to bridge this gap. Our work combines the use of these new platforms with co-creation and collaboration tools for AI assisted remote Work-Integrated-Learning (WIL) opportunities, including efforts in community and with the public library system.  This paper outlines some of our experiences to date, and proposes methods to further integrate  AI education into community-driven applications for remote WIL.}

\end{quote}
\end{abstract}

\section{Background and Introduction}
\begin{figure*}[h!]
  \centering
  \includegraphics[scale=0.61] {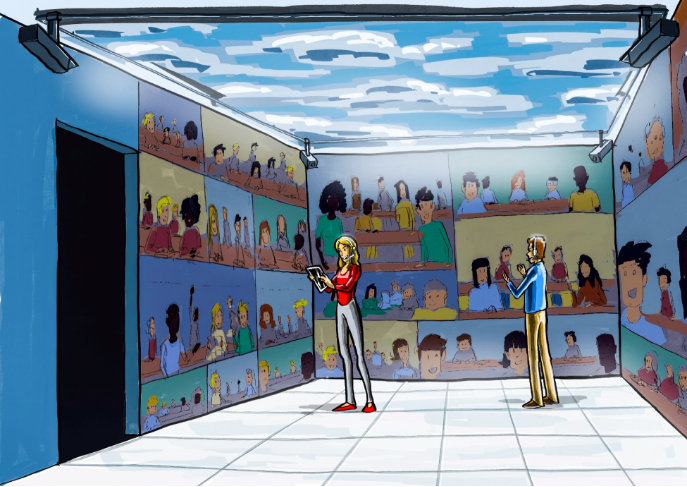}
  \includegraphics[scale=0.15] {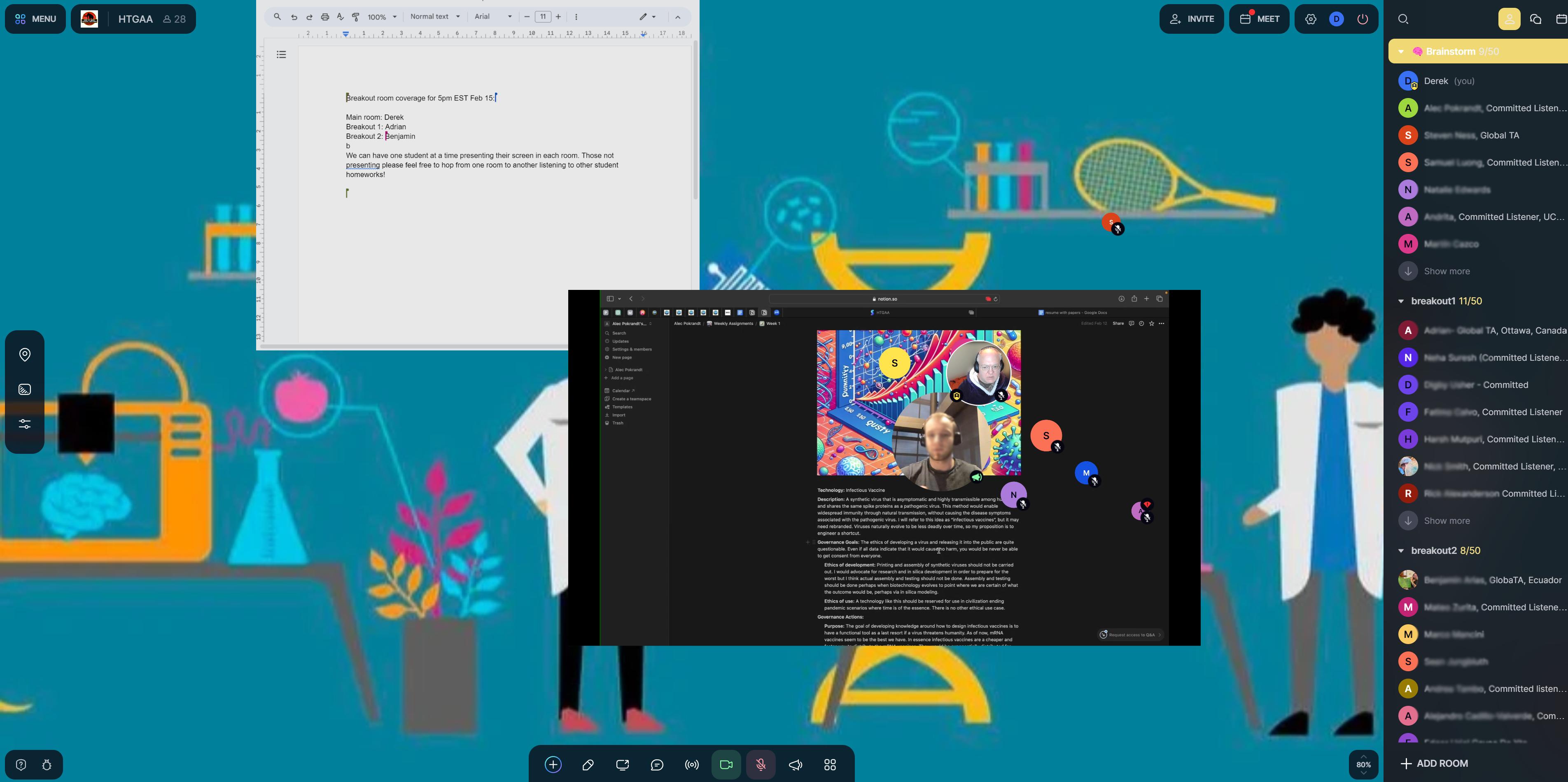}
  \caption{   Idealized team-based collaboration with remote participants in cohorts (left), and a {\em Spatial Chat} environment for individual remote participants (right).}
  \label{fig:platforms}
\end{figure*}

 Efforts like the Connected Coast\footnote{  Connected Coast Project https://connectedcoast.ca/} and satellite-based broadband are crucial in tackling the internet connectivity challenges in remote areas. The Connected Coast extends to individuals living in 139 rural and remote communities, including 48 Indigenous communities representing 44 First Nations. 
 
 As shown in Figure~\ref{fig:platforms}, this infrastructure can now be leveraged in more effective remote learning platforms, allowing team-based activities that include participants who otherwise cannot collaborate in-person. On the left, the Figure illustrates course instructors as facilitators of a variety of remote cohorts interacting with each other---and possibly AI avatars!  This environment is explicitly team-based and student-led.  On the right, learners collaborating as individual participants, where teams can dynamically be formed by ``moving" to an area of the shared space~\footnote{  Spatial Chat https://www.spatial.chat/}.  A key component within these new platforms is providing cloud-based computing resources, such as cloud-based virtual machines, to remote learners, addressing the lack of access to physical computing hardware.
 

Increasing diversity in AI education 
requires innovative methods to face the challenges in the realm of AI learning, especially in remote communities with non-traditional learners. Guided by the Four Rs---Respect, Relevance, Reciprocity, and Responsibility~\cite{Kirkness1991FirstNA}---along with insights on online learning practices that are culturally responsive, interactive and project-based~\cite{Hunt2020OnlineLP}, WIL efforts for inclusive AI education is a promising way to reach remote, non-traditional learners and build a more just and equitable future for all.

Of course, these tools and the approach to using them must be developed in collaboration with the communities that the experiences are designed for. We describe a series of efforts to achieve cultural integration of AI education through the use of participatory design approaches.

The remainder of this extended abstract is organized as follows.  First, we consider lessons learned through reciprocal learning from WIL in a Canadian context.  We then take inspiration from remote education in the context of a global course. 
 Next, we consider participatory design approaches to developing community classes with experience in the USA and Mexico. Finally, issues when  designing for Indigenous and rural communities, and the role of AI and avatars in education are considered, before a brief discussion, conclusions and next steps.

\section{Lessons Learned from WIL}

With support from CEWIL Canada\footnote{  Work Integrated Learning https://cewilcanada.ca/}, we were able to explore both in-person and online WIL opportunities with Indigenous learners in remote communities. In each case, our reciprocal lessons learned inform us about how AI education was received, and plant the seeds as to how we can improve our methods in the future. 

\subsubsection{The Uchucklesaht Tribe\footnote{  Uchucklesaht Tribe https://www.uchucklesaht.ca/}}
Our initial work involved direct, in-person engagement with the Uchucklesaht community. We had a class of fourth-year distributed systems students focus on AI-driven geospatial analytic tools for kelp farm site planning, which was the area of focus for the community at the time. The full class of students were given remote access to community elders and provided cultural awareness training. There were some successes with this project, however, the approach had limitations. It became evident that the deep understanding of AI techniques and software modification skills were not effectively transferred to the community members, and the overall impact of the educational opportunities was much more transient that we had hoped. This challenge highlighted the need for more sustainable and accessible educational models in AI.


\subsubsection{The Indigenous Matriarchs Four\footnote{  IM4 Lab https://im4lab.com/im4/} Virtual Production Project}
We recently provided some technical assistance in a project in Virtual Production for remote Indigenous filmmakers which introduced sophisticated AI tools to learners without a technical background. These tools included motion capture models and generative AI for asset creation. However, the challenge persisted in making these tools accessible and understandable to learners, especially those who were new to the technology involved. These were highly motivated learners who had deep needs to apply the tools to advance their art and careers, but the software and the computing background that it required got in the way of learning. This emphasized the need for AI education to be more intuitive and user-friendly. 

A key component of our approach in this project was providing cloud-based computing resources, such as AWS virtual machines, to remote learners. This addressed the lack of access to physical computing hardware. The format of the remote learning experience, presented on zoom and using breakout rooms for smaller groups, was effective but demonstrated problems with this community in terms of encouraging participation from those individuals still building their technical confidence. The Indigenous Matriarchs 4 lab were able to establish a fantastic cultural rapport with the students that helped to overcome these issues, but it was clear that the tools themselves were insufficient in some important ways.

\subsubsection{Community WIL}
Work-Integrated-Learning is most often an industry-led effort, at least in the formal co-ops and internships that have historically dominated the understanding. We have been developing collaborations with community organizations such as makerspaces\footnote{  Victoria Makerspace https://makerspace.ca/}, makerspace events\footnote{  Makers Making Change https://www.makersmakingchange.com/s/}, and non-profits\footnote{  Pacific Coastal Computing Association (PCCA) https://pacificcoastalcomputing.ca/}  which allow us greater integration of academic topics and student productivity towards very practical projects. These more application-oriented activities are very motivating for students and take advantage of some of the natural affordances of community workshop spaces. A more formal investigation of community workshops and the equity advantages of assistive aids can be found in a recent ACM SIGACCESS paper \cite{higgins_towards_2023} which describes the interaction with purely university-focused makerspaces. Once the makerspace efforts reach further away from academia the advantages are even more pronounced. 

In our own particular set of efforts, graduate students from Northeastern University are working on extensions to the current Spatial Chat platform and the creation of speech-driven avatars for users with autism-related speech production difficulties. The placement in a makerspace with the capability to develop accelerated Surface Mount Technology (SMT) electronics has the potential to facilitate the creation of hardware solutions that use AI as assistive devices that are very highly personalized to the individual being assisted.  In British Columbia, this development of hardware devices is an important alternative to a cell phone app.  Not only because of the hardware acceleration for AI, but also because of new restrictive policies on cell phone use in schools \cite{press__bc_2024}. 

This set of equity-focused academic activities in combination with community institutions that focus on practical applications of computing for disabilities improves two things: (1) our students' understanding of computing and it's impact on people with disabilities, and (2) the lives of the individuals included in the projects.


\section{Remote Learning and Inclusion in a Global Remote Class}
The lead author on this paper has been involved with the synthetic biology community since 2008 when he formed a student team to compete in the International Genetically Engineered Machines competition (iGEM) \cite{vilanova_igem_2014}. This synthetic biology community is an interesting one from the perspective of AI education in groups that are not traditionally a core AI audience. In recent years the role of AI in biology has become indisputable with highly successful efforts on hard problems such as protein folding \cite{varadi_alphafold_2024}, but for most early-career biologists, becoming an AI expert is still not an obvious training path. In the synthetic biology community in particular, there are grassroots efforts to encourage diversity in forms that are not yet common in computer science. 

Open source hardware design efforts (such as those in microfluidic devices \cite{kong_open-source_2017}) have obvious analogues in open software repositories, but the focus on interdisciplinary teams and production of new lifeforms means that synthetic biology has developed interesting ethical and team interaction structures that have not yet been as obvious in core computer science disciplines \cite{santolini_igem_2023}. 

One of the most inclusive science outreach efforts in this space has been the community focus of the MIT Media Lab, both in a yearly conference and a global course called {\em How To Grow Almost Anything} (HTGAA\footnote{  HTGAA https://htgaa.org/}) \cite{perry_how_2022}. A notable element of the HTGAA course has been the community organizing structures in place to support remote students. Developed in part through the pandemic when working in physical labs with others was not safe, the course has developed an emphasis on robotic lab support systems and this year has been supported by large equipment and service donations that have allowed global community labs and over 500 students around the world to participate in the wetlab activities remotely, hosting the lab robots in these worldwide community labs. The students are supported by a team of over 40 teaching assistants from past iterations of the course, using the Spatial Chat platform described in this paper. 

In the yearly Community Biology Summit\footnote{  GCBS https://www.biosummit.org/about} one of the organizing innovations has been to use corporate sponsorship to pay for travel and accommodation costs for participants from around the world. Not all participants receive such support, but a committee of organizers reviews applications from students in financial need and the inclusivity of the community has been greatly strengthened by amplification of these voices. Not all of these efforts are directly around AI education, of course, but the design of synthetic biology constructs are being increasingly intermediated by AI systems and the training of subject matter experts in building AI driven applications is becoming ever more relevant. It has not escaped the authors' notice that these efforts in biology often result in a gender balance of trainees that are the mirror image of many efforts in traditional AI courses, for instance 89\% female in one University of Wisconsin study \cite{dutton_novel_2023}.




\section*{Methodological Approach to Community Engagement and Skill Development}
\begin{figure}[h!]
  \centering
  \includegraphics[scale=0.2] {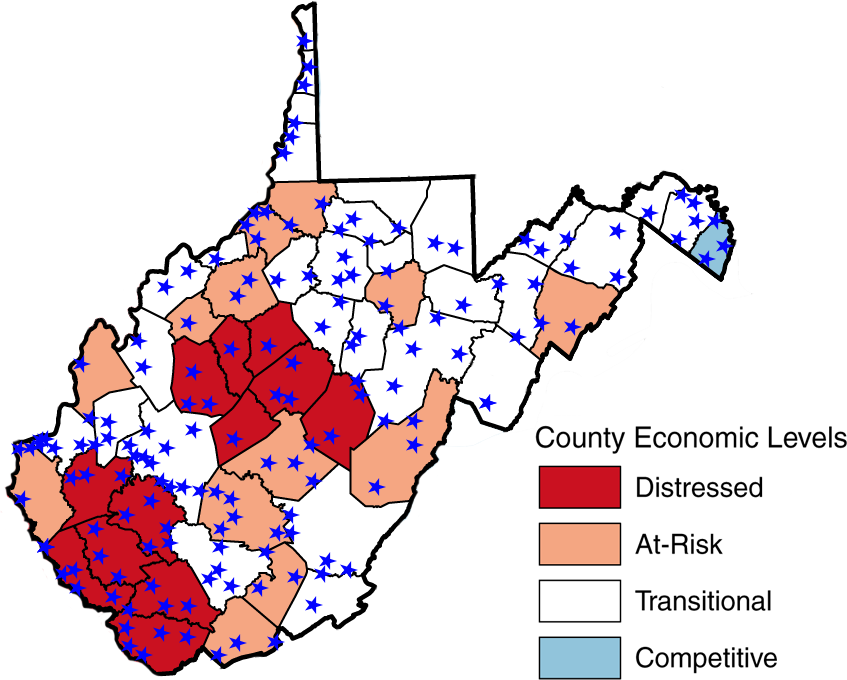}
  \caption{   Leveraging Public Libraries to Empower Economic Development in Rural Areas: This map, marked with blue stars to denote public libraries, overlays the economic distress zones in West Virginia. Our initiative utilizes the public libraries in the U.S. as vital hubs for conducting in-person studies, where we engage with local communities to teach adult workers essential computer skills, enhancing their employability.}
  \label{fig:libraries}
\end{figure}
Our methodological framework was meticulously crafted to address the unique educational and skill development needs of rural and indigenous communities through the innovative application of AI-enhanced Work-Integrated Learning (WIL). This strategic approach was anchored in three cornerstone activities, each carefully designed to foster an environment of learning and growth tailored to the specific contexts of these communities. The activities were as follows: establishing \textit{Initial Skill Goals} in close collaboration with community members, launching a \textit{First Iteration of Classes} to gauge and meet the immediate learning needs, and implementing a \textit{Second Iteration of Classes} to refine and expand upon the lessons learned in the initial phase. Fig. \ref{fig:libraries}
shows how we levarage public libraries to support our efforts.

The first of these activities, defining \textit{Initial Skill Goals}, involved an in-depth dialogue with community members to identify the skills most relevant and beneficial to their socio-economic advancement. This phase was crucial for ensuring that our educational initiatives were not only aligned with the community's aspirations but also responsive to the local job market's demands. By engaging directly with a broad cross-section of the community, including educators, local leaders, and potential learners, we were able to craft a curriculum that was both ambitious and achievable, laying a solid foundation for the subsequent phases of the project.

Following this initial groundwork, we embarked on the \textit{First Iteration of Classes}. This phase was pivotal for testing our hypotheses about the community's learning needs and preferences in a real-world setting. The classes served as a platform for interactive learning and provided us with invaluable feedback on the curriculum's relevance and effectiveness. This iterative process was not just about imparting knowledge but also about creating a dynamic learning environment where feedback from participants could immediately inform improvements and adjustments to the program.

Building on the insights gained from the first iteration, we proceeded to the \textit{Second Iteration of Classes}, which aimed to refine our approach based on the participants' feedback and the initial outcomes. This phase allowed us to tailor the learning experience more closely to the participants' evolving needs, introducing more nuanced and advanced topics as required. By adopting a flexible and responsive teaching methodology, we ensured that the learning experience remained relevant, engaging, and impactful for all participants. Our overarching goal throughout this methodological journey was to deliver immediate and tangible benefits to the participants, reflecting our commitment to an ethical research paradigm that values the contributions and welfare of community members. This approach not only provided a solid foundation for our research but also fostered a sense of ownership and engagement among the community members, ensuring that the learning solutions we developed were genuinely rooted in the community's needs and aspirations. Through this collaborative and adaptive process, we aimed to not only enhance the educational landscape for rural and indigenous communities but also to empower these communities to shape their own futures through the transformative power of AI-enhanced learning.

\subsubsection*{Initial Skill Goals}
The initial phase of our project involved engaging with approximately 20 members of the rural and indigenous communities we aimed to serve. This engagement was critical in laying a robust foundation for our educational interventions, with the primary objective of delineating the specific skill sets required and desired by the community in alignment with the demands of the local job market. To achieve this, we embarked on a comprehensive consultation process, engaging in detailed discussions with key local stakeholders who play pivotal roles in education and employment within these communities. These stakeholders included the board directors of local libraries, who are often at the forefront of community education initiatives; professors and instructors from community colleges, who have a direct understanding of the educational needs and opportunities; and high school guidance counselors, who provide career guidance to the youth and have insights into the transition from education to employment.

Supplementing these discussions, we conducted a thorough analysis of current job postings in the local area. This analysis served a dual purpose: firstly, to validate and reinforce the findings from our stakeholder discussions regarding the demand for specific skills, and secondly, to ensure that the educational content we planned to develop would have practical utility and relevance in the job market, thereby enhancing employment opportunities for community members. The feedback we received from these engagements was remarkably consistent, highlighting a widespread demand for basic "computer skills" as a fundamental prerequisite for a broad range of employment opportunities. Upon further investigation, we discovered that this term—often used broadly in job descriptions—specifically referred to the ability to competently navigate the Windows Operating System and to possess a foundational understanding of Microsoft Office applications, including Word, Excel, and PowerPoint. These skills were not merely desirable but were considered essential for job seekers in the community to be competitive in the local job market.

Armed with this critical insight, we tailored our initial set of educational offerings to focus squarely on these ``computer skills.'' Recognizing the foundational importance of these skills for employment, we designed our classes to provide hands-on, practical training in the Windows Operating System and Microsoft Office suite. Our goal was to ensure that participants not only understood the theoretical aspects of these tools but could also apply them effectively in real-world scenarios, thereby significantly enhancing their employability and confidence in navigating the digital aspects of the modern workplace.

This focus on practical, applicable skills reflects our commitment to delivering educational programs that are directly relevant to the needs and aspirations of the community members we serve. By starting with a foundational layer of digital literacy, we aim to build a platform upon which participants can further develop their skills, explore new opportunities, and ultimately achieve greater success in the job market.

\subsubsection*{First Iteration of Classes}

The inaugural class attracted five participants, serving as a practical assessment of the community's skill needs. Despite planning an overview of the Office suite, the session was primarily driven by participant inquiries, revealing a significant underestimation of the participants' starting skill levels. This session, originally intended to cover a broader range of content, was adjusted to focus solely on basic functionalities within Microsoft Word. The engagement served as a valuable data collection method, surpassing traditional approaches like focus groups in understanding the community's learning requirements.

\subsubsection*{Second Iteration of Classes}

Adjusting our strategy based on the insights gained, the second set of classes was designed to cater to a more basic starting skill level, spanning an entire day with sessions lasting an hour each. However, the turnout and diverse skill levels of the four participants challenged our curriculum's structure, underscoring the limited pool of participants in smaller communities and the difficulty in catering to varied skill levels within a single session. This iteration highlighted the necessity of flexible and adaptive teaching strategies to accommodate the diverse learning paces and needs of participants.
\begin{figure}[h!]
  \centering
  \includegraphics[scale=0.38] {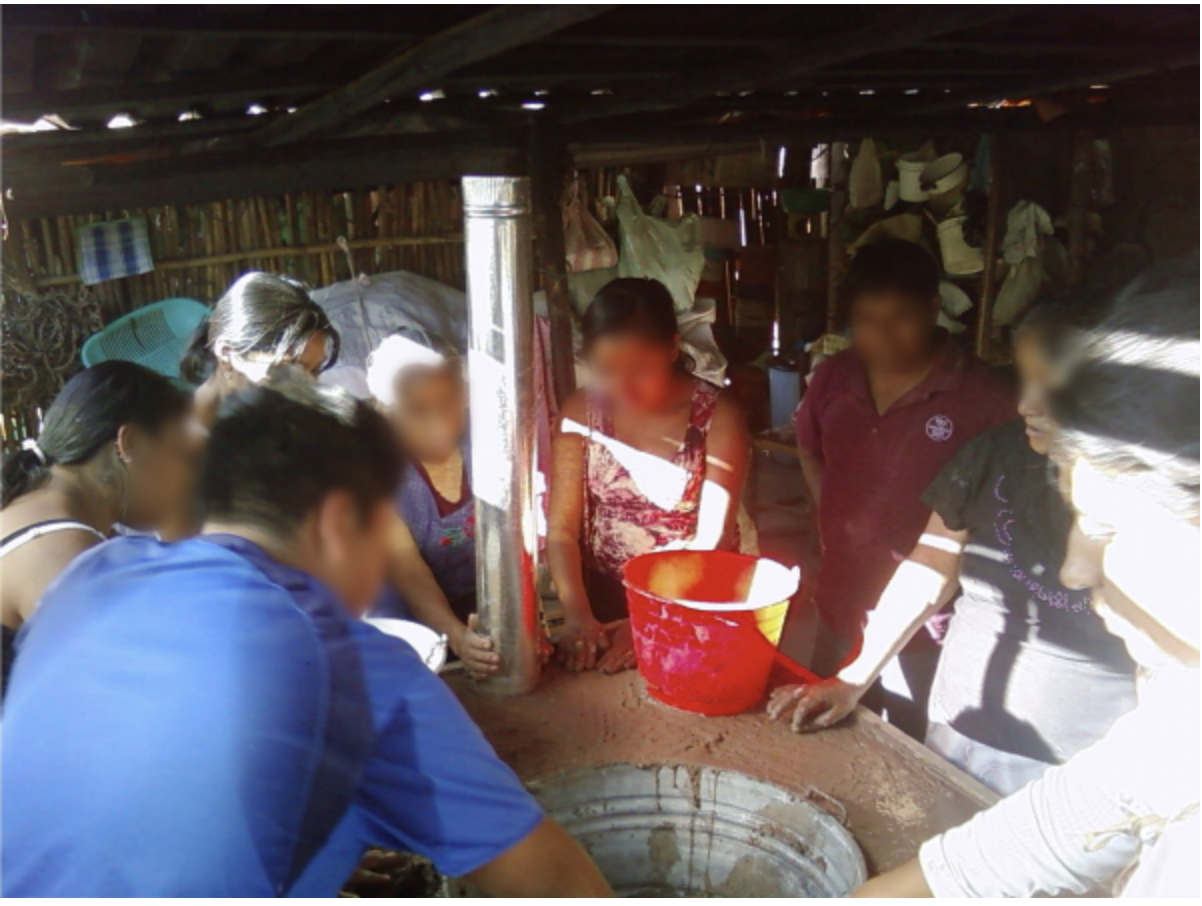}
  \caption{   Fostering Cultural Integration Through Participatory Design: This photograph captures the essence of our collaborative sessions, where indigenous and rural communities actively contribute their cultural heritage and traditions into the co-creation of AI-enhanced technologies. It illustrates our commitment to working alongside these communities, empowering them to take the lead in designing and developing technological solutions that resonate with their cultural identity and meet their unique needs.}
  \label{fig:indigenous}
\end{figure}
\subsubsection*{Integrating Cultural Theory and Participatory Design}

To further enrich our methodological framework, we integrated cultural theory, particularly Hofstede's cultural dimensions, to tailor our educational tools to the community's cultural values and learning preferences. Moreover, we incorporated participatory design sessions, empowering community members to actively contribute to the technology creation process. These sessions were envisaged as interactive workshops where participants could express their preferences, engage in prototype design, and directly influence the development of AI tools. This approach aimed to democratize technology creation, ensuring the tools developed were not only functional but also culturally resonant and community-driven. Figure \ref{fig:indigenous} showcases examples of our participatory design sessions in Indigenous rural communities in Guerrero, Mexico. 

Through these methodical steps, our research endeavors to bridge the educational gap for rural and indigenous communities, leveraging AI and participatory design to create culturally sensitive, accessible, and effective educational experiences. This holistic approach emphasizes the importance of community engagement, cultural understanding, and adaptive learning strategies in the development of educational technologies.

\section*{Designing Geospatial Applications for Indigenous and Rural Communities} To effectively support the educational and developmental needs of rural and indigenous communities in Mexico, Canada, and the US, our approach to interface design is centered around AI-enhanced Work-Integrated Learning (WIL) activities. These interfaces will prioritize practical, application-driven AI tools that are deeply rooted in the local contexts, enabling users to engage critically with AI technology, especially in identifying and mitigating biases. A pivotal aspect of our strategy involves the innovative employment of AI-driven personas to facilitate users' understanding and navigation of potential biases within language models. This method not only highlights the limitations inherent in current AI guardrails but also promotes a comprehensive appreciation of diverse perspectives, essential in these culturally rich environments.

To bridge the gap between technology and cultural understanding, we will integrate cultural theory into our design process, specifically drawing upon Hofstede's cultural dimensions theory. This framework will guide us in tailoring the interfaces to reflect the cultural values, communication styles, and learning preferences of each community, ensuring that the technology is culturally relevant and sensitive. By doing so, we aim to create interfaces that are not just tools but are extensions of the community's cultural fabric, enhancing the educational experience through a culturally informed lens.

Moreover, the design of these interfaces will incorporate participatory design sessions, actively involving community members in the creation process. These sessions will be structured to empower communities, enabling them to become co-creators of technology rather than mere consumers. Through workshops and collaborative design activities, participants will have the opportunity to express their needs, preferences, and cultural insights, which will be instrumental in shaping the development of the interfaces. This participatory approach will foster a sense of ownership and relevance among the users, ensuring that the final product genuinely resonates with their educational and cultural realities.

The participatory design sessions will be characterized by interactive, hands-on activities that encourage creativity and open dialogue. Participants will engage in scenario building, sketching interfaces, and prototyping, guided by facilitators who are sensitive to cultural nuances and adept at bridging technological concepts with local knowledge. By involving the communities in every step of the design process, we aim to democratize technology creation, enabling indigenous and rural communities to tailor AI tools to their unique cultural contexts and learning environments.

This comprehensive approach to interface design, which marries advanced AI capabilities with a profound respect for cultural diversity and participatory methodologies, promises to create learning environments that are not only technologically innovative but also deeply embedded in the cultural and social tapestry of indigenous and rural communities. Through this fusion of technology and culture, we aspire to empower these communities, transforming them from passive recipients of technology to active creators and shapers of their educational tools.

\section{AI Avatars in Remote Education}
In particular for AI-avatar and communication platform driven efforts, there is no need to restrict ourselves to local WIL experiences. We propose structuring remote WIL activities to demonstrate how and when to use AI, from a practical, application driven, perspective, incorporating critical assessment as part of its use, from two perspectives: (1) identifying and overcoming bias, and (2) personalizing non-hierarchical learning environments.

\subsubsection{Overcoming Bias} Ideally, automated personas driven by AI could be a natural way of assessing bias in language models and help learners understand the impact of model guardrails on creating diverse personas for interaction.  In preliminary work, we have been exploring how current guardrail approaches currently constrain the experiences of those outside the norm.  From both a content-generation perspective, and a diversity of dialogue perspective, students will find a lack of flexibility in current AI alignment approaches that is detrimental to the type of interactions that they may want to encourage. For example, in an effort associated with the Climate Disaster Project\footnote{  Climate Disaster Project https://climatedisasterproject.com/} we have been attempting to create an AI persona of a ``climate change denier" to allow students to practice advocating their truths to someone with opposing viewpoints. An exciting result is that this is virtually impossible with the guardrails on many commercial large language models~\cite{badyal}.

\subsubsection{Personalizing Non-hierarchical Learning Models with Conversational AI}
Once understanding of bias has been established, students can explore the ways in which conversational AI models could transform the traditional, hierarchical classroom experiences.  To this end, a more inclusive and student-led learning environment would include allowing learners (who understand the limitations of AI) to direct conversations with AI personas. Rather than a single instructor-led classroom presentation, or a set of fixed breakout rooms, this would include a spatial chat interface to allow students to flexibly self-assemble into smaller groups---including AI personas---while maintaining awareness of what is going on with the rest of the class. We've experimented with this environment in such events as student-led poster sessions and had promising results.

\section*{Discussion}
There are two future directions we would like to specifically highlight from a cultural perspective with these communities, but which ultimately apply to all learners. First is the use of AI personas/avatars to explore the capability of the technology involved, understanding where bias is introduced.  Second is to reduce the social hierarchy distance between learners and teachers to allow everyone to  freely ask questions and converse without constraint in naturally formed teams. Learners can themselves decide who they will co-locate with---including an AI driven persona. The spatial chat mechanism allows students to control who can see and hear them by remaining in control of their proximity to others.

Our experiences and proposed future methods in AI education for remote communities underscore the importance of evolving and adapting teaching methodologies to address digital inequality. Our journey highlights the challenges and successes in making exposure to AI education inclusive and accessible. As we continue to innovate and implement AI educational tools, our focus remains on creating equitable learning opportunities for students from all backgrounds, particularly those in remote and under-served communities.

\subsection*{Reflections for the Future} Reflecting on our most recent project, which focused on harnessing AI-enhanced Work-Integrated Learning (WIL) to meet the educational and developmental needs of rural and indigenous communities \cite{chun2019designing,flores2020challenges,hanrahan2020reciprocal,angel2015participatory,chiang2018exploring}, we revisit the traditional methodologies employed in design research, such as focus groups and interviews. These methods have long been the cornerstone for grounding the development of technological solutions in the rich, nuanced data derived from real-world contexts. By engaging directly with the communities we aim to serve, we gather invaluable insights into their unique challenges, aspirations, and the cultural intricacies that shape their interaction with technology. This foundational approach ensures that our design process is deeply informed by the lived experiences of our target users.

However, our project took a distinctive path by emphasizing the immediate impact and tangible benefits for our participants right from the outset. Unlike conventional approaches that primarily focus on the technological output—the final artifact—our strategy is deeply rooted in achieving meaningful human outcomes. This shift towards prioritizing the well-being and empowerment of community members reflects a more humane and ethical approach to technology design. It acknowledges the participants not merely as subjects of research but as active stakeholders whose immediate needs and long-term aspirations are central to the design process.

This reorientation towards creating immediate value represents a significant departure from traditional technology-first perspectives. It challenges us to think creatively about how our interventions can provide immediate relief, enhancement, or empowerment to individuals and communities. By doing so, we ensure that our research is not just a theoretical exercise but a practical endeavor that has a direct and positive impact on the lives of those involved. This approach has the dual benefit of making our research more ethically responsible and ensuring that the solutions we develop are genuinely aligned with the needs, contexts, and cultures of the communities we aim to support.

Moreover, by integrating this focus on immediate benefits, our project aligns more closely with the principles of participatory design, where the design process is collaborative and inclusive, allowing participants to have a say in the creation of solutions that affect their lives. This ensures that the technology we develop is not only functional and efficient but also culturally sensitive, accessible, and relevant to the people it is intended to help. In this way, our work not only contributes to the academic and practical fields of AI and WIL but also stands as a testament to the importance of ethical, human-centered design in addressing the challenges faced by rural and indigenous communities.

In our methodological approach, we engaged in activities such as running community classes, which served dual purposes: gathering data about community needs and providing direct value through educational opportunities. This approach, akin to participatory design, allowed us to prototype micro-solutions that offered immediate benefits to participants, thereby avoiding the common pitfall of research fatigue. Our findings suggest that such micro-activities, whether they be educational classes, community advocacy, or serving as a liaison, can transform data gathering into meaningful engagements that provide both immediate benefits to participants and insightful proofs-of-concept for design research.

Moreover, our collaboration with local organizations like rural libraries highlighted unexpected benefits, as these institutions found value in using our courses to attract attendees and enhance their programming. This reciprocal value exchange underscores the importance of delivering immediate benefits through research activities, aligning with citizen science principles where participants contribute to scientific research while gaining personal benefits, such as increased scientific literacy. By integrating citizen science techniques and insights from computing for social good, we advocate for design research that delivers tangible value to participants, particularly in rural communities where engagement fatigue is a real concern. Our approach demonstrates the potential for research activities to offer immediate results and value, fostering a more inclusive and participatory design process. This not only benefits the communities involved but also enriches the research with grounded data and experiences, paving the way for more meaningful and impactful technological interventions.





\section*{Conclusion and Next Steps}
In conclusion, our dedicated efforts to cultivate diversity within educational settings, especially targeting non-traditional learners in remote and underserved communities, are met with formidable obstacles. Among these, the most pressing is the acute shortage of resources that are readily accessible to these learners, a challenge that has historically hindered equitable educational opportunities. However, the emergence of platforms designed for remote and hybrid learning in the aftermath of the global pandemic offers a beacon of hope. These platforms, with their inherent capacity for facilitating online team-based collaboration, provide a unique opportunity to bridge the gap in educational access and quality that exists for remote learners.

Our methodology, which harmonizes the advanced functionalities of these platforms with the dynamic potential of co-creation and collaboration tools, is designed to enhance Work-Integrated Learning (WIL) opportunities that are assisted by artificial intelligence. This blend of technology and pedagogy aims not just to supplement traditional learning models but to revolutionize them, making learning more adaptable, interactive, and engaging for learners outside the conventional classroom setting.

Through the insights and experiences shared in this extended abstract, we have outlined our journey and the strategies we have employed to integrate AI education more deeply into community-driven applications, specifically tailored for remote WIL. Our vision extends beyond the mere adoption of new technologies; we aim to fundamentally transform the landscape of education to be more inclusive, responsive, and accessible to learners from all walks of life. By delving deeper into the exploration and refinement of these innovative methods, we are committed to unlocking the untapped potential of remote learning environments. Our goal is to dismantle the barriers that restrict educational access, thereby democratizing learning and empowering individuals across diverse geographical, cultural, and socio-economic backgrounds.

This endeavor is not without its challenges, but the preliminary successes and the positive feedback from our community engagements have fortified our resolve to continue this work. As we progress, we remain open to learning, adapting, and evolving our strategies to meet the needs of our learners more effectively. In doing so, we aspire to create a future where quality education is not a privilege confined to specific locales or demographics but a universal right that transcends boundaries and barriers. By fostering an educational ecosystem that is more equitable, we pave the way for a more inclusive and diverse global community, empowered by knowledge and the endless possibilities that it brings.

\section{Acknowledgements}  This work was partially supported by the NSF grant FW-HTF-19541.
\bibliographystyle{aaai24}
\bibliography{bibfile}

\end{document}